\RequirePackage{lineno}
\documentclass[twocolumn,aps,prd,amsmath,amssymb,floatfix,showkeys,nofootinbib,superscriptaddress,showpacs]{revtex4}

\usepackage{graphicx}
\usepackage{optional}
\usepackage{url}

\def\units#1{\hbox{$\,{\rm #1}$}}
\def\degrees{\hbox{${}^\circ$}}

\usepackage{amsthm}
\usepackage{wasysym}
\usepackage{graphicx}
\usepackage{array}

\newcolumntype{L}[1]{>{\raggedright\let\newline\\\arraybackslash\hspace{0pt}}m{#1}}
\newcolumntype{C}[1]{>{\centering\let\newline\\\arraybackslash\hspace{0pt}}m{#1}}
\newcolumntype{R}[1]{>{\raggedleft\let\newline\\\arraybackslash\hspace{0pt}}m{#1}}

\begin{document}

\title{Measurement of the high-energy gamma-ray emission from the Moon with the Fermi Large Area Telescope\\Author list created Monday 19 Oct 2015 06:39 PDT}
\author{M.~Ackermann}
\affiliation{Deutsches Elektronen Synchrotron DESY, D-15738 Zeuthen, Germany}
\author{M.~Ajello}
\affiliation{Department of Physics and Astronomy, Clemson University, Kinard Lab of Physics, Clemson, SC 29634-0978, USA}
\author{A.~Albert}
\affiliation{W. W. Hansen Experimental Physics Laboratory, Kavli Institute for Particle Astrophysics and Cosmology, Department of Physics and SLAC National Accelerator Laboratory, Stanford University, Stanford, CA 94305, USA}
\author{W.~B.~Atwood}
\affiliation{Santa Cruz Institute for Particle Physics, Department of Physics and Department of Astronomy and Astrophysics, University of California at Santa Cruz, Santa Cruz, CA 95064, USA}
\author{L.~Baldini}
\affiliation{Universit\`a di Pisa and Istituto Nazionale di Fisica Nucleare, Sezione di Pisa I-56127 Pisa, Italy}
\affiliation{W. W. Hansen Experimental Physics Laboratory, Kavli Institute for Particle Astrophysics and Cosmology, Department of Physics and SLAC National Accelerator Laboratory, Stanford University, Stanford, CA 94305, USA}
\author{G.~Barbiellini}
\affiliation{Istituto Nazionale di Fisica Nucleare, Sezione di Trieste, I-34127 Trieste, Italy}
\affiliation{Dipartimento di Fisica, Universit\`a di Trieste, I-34127 Trieste, Italy}
\author{D.~Bastieri}
\affiliation{Istituto Nazionale di Fisica Nucleare, Sezione di Padova, I-35131 Padova, Italy}
\affiliation{Dipartimento di Fisica e Astronomia ``G. Galilei'', Universit\`a di Padova, I-35131 Padova, Italy}
\author{R.~Bellazzini}
\affiliation{Istituto Nazionale di Fisica Nucleare, Sezione di Pisa, I-56127 Pisa, Italy}
\author{E.~Bissaldi}
\affiliation{Istituto Nazionale di Fisica Nucleare, Sezione di Bari, I-70126 Bari, Italy}
\author{R.~D.~Blandford}
\affiliation{W. W. Hansen Experimental Physics Laboratory, Kavli Institute for Particle Astrophysics and Cosmology, Department of Physics and SLAC National Accelerator Laboratory, Stanford University, Stanford, CA 94305, USA}
% ...

\collaboration{The Fermi LAT Collaboration} \noaffiliation

\author{F.~Cerutti}
\affiliation{European Organization for Nuclear Research (CERN), CH-1211 Geneva, Switzerland}
\author{A.~Ferrari}
\affiliation{European Organization for Nuclear Research (CERN), CH-1211 Geneva, Switzerland}
\author{P.~R.~Sala}
\affiliation{Istituto Nazionale di Fisica Nucleare, Sezione di Milano, I-20133 Milano, Italy}

\title{Measurement of the high-energy gamma-ray emission from the Moon with the Fermi Large Area Telescope}

\date{today}

\begin{abstract}

We have measured the gamma-ray emission spectrum of the Moon using 
the data collected by the Large Area Telescope onboard the Fermi
satellite during its first $7$ years of operation, in the energy
range from $30 \units{MeV}$ up to a few $\units{GeV}$.
We have also studied the time evolution of the flux, finding 
a correlation with the solar activity.
We have developed a full Monte Carlo simulation describing
the interactions of cosmic rays with the lunar surface. The results of the 
present analysis can be explained in the framework of this model,
where the production of gamma rays is due to the interactions 
of cosmic-ray proton and helium nuclei with the surface of the Moon. 
Finally, we have used our simulation to derive the cosmic-ray proton
and helium spectra near Earth from the Moon gamma-ray data.

\end{abstract}


\maketitle


\section{Introduction}
\label{sec:intro}

The Moon, as well as other bodies in the solar system, 
can be passive sources of high-energy gamma rays, resulting from inelastic collisions 
of energetic cosmic-ray (CR) particles with their material~\cite{morris}.  
A measurement of the lunar gamma-ray flux therefore represents a useful tool to 
investigate the properties of CRs outside Earth's magnetic field.   
Such a study does require accurate modeling of the interaction processes 
of high-energy CRs with the lunar surface. 

The emission of high-energy gamma rays from the Moon was first observed
by the EGRET experiment~\cite{egret}, which operated from 1991 to 2000
on the Compton Gamma Ray Observatory (CGRO). More precise results were 
recently published by the Fermi LAT Collaboration using the data
collected by the Large Area Telescope (LAT) during its first $2$ 
years of operation~\cite{abdo2012}, which provided a measurement of the
gamma-ray flux above $100\units{MeV}$. 

In the present work we have evaluated the gamma-ray flux from the 
Moon using the data collected by the Fermi LAT in its first $7$ years of
operation, from August 2008 to June 2015. 
Not only is the current data set much larger, but
the data were processed with the newest {\tt Pass 8} reconstruction 
and event-level analysis ~\cite{atwood2012}, allowing  the useful energy range 
to be extended well below $100\units{MeV}$. We have studied the 
time evolution of the gamma-ray flux from the Moon, finding the 
expected correlation with the solar activity.

Gamma rays from the Moon are mainly emitted with sub-$\units{GeV}$ 
energies and their flux depends on the fluxes of CRs impinging 
on the Moon and on their inelastic interactions with the lunar regolith. 
The chemical composition of the lunar surface  
also plays a crucial role in determining the gamma-ray yield. 
As will be discussed in sec.~\ref{sec:fluka}, the
energy spectrum of lunar gamma rays is sensitive 
to the spectra of CR primaries in the range up to a few 
tens of $\units{GeV/n}$, which are strongly affected by 
the solar activity. 

Therefore, the main ingredients of any model aiming to 
provide an interpretation of the gamma-ray emission from the Moon are: 
(a) the interactions of CRs with matter; 
(b) the lunar surface composition; 
(c) the CR energy spectra.   
The models describing inelastic interactions of CRs with matter
are well validated against the data from accelerator experiments
and are quite reliable in the energy range of interest.
The predicted gamma-ray spectra will therefore depend on 
the input CR spectra and on the lunar surface composition. 

Simultaneous measurements of the lunar gamma-ray spectrum
and of the spectra of charged CRs near Earth can provide 
the possibility to test the chemical composition of the 
lunar surface. In fact, the CR energy spectra provided as input to 
the models are usually evaluated from the data collected 
in a different epoch and accounting for solar modulation.
The simultaneity allows eliminating uncertainties on
the CR spectra due to solar modulation.
The AMS-02 instrument is currently taking data simultaneously with 
the Fermi LAT, and recently its measurements of the CR proton and 
helium energy spectra near Earth have been published~\cite{ams02,amshe}. 
This fact therefore offers the unprecedented possibility 
to set severe constraints on the lunar gamma-ray emission models. 

In this work we have implemented a full Monte Carlo simulation of the 
CR interactions with the Moon surface based on the 
FLUKA~\cite{battistoni2007,ferrari2005,flukaweb} code. 
In our simulation we assumed a lunar surface
chemical composition derived from the samples of lunar rock 
taken by the astronauts of the Apollo missions~\cite{apollo}.
We show that the simulation reproduces accurately the 
Moon gamma-ray data taken by the LAT in the same epoch as the AMS-02 
proton and helium data. Finally, starting from a model 
of the local interstellar spectra (LIS) of CR protons and helium nuclei, 
we have fitted the Moon gamma-ray data using the gamma-ray
yields predicted by our simulation to derive the CR proton and helium 
spectra at $1 \units{A.U.}$ from the Sun 
and to evaluate solar modulation potential.

\section{The lunar gamma-ray emission spectrum}
\label{sec:formulae}

As mentioned in sec.~\ref{sec:intro}, gamma rays emitted from the Moon
are produced after inelastic interactions of charged CRs with the 
lunar surface. Hereafter we will make the assumption 
that the CR flux on the lunar surface is spatially isotropic.

Indicating with $I_{i}(T)$ the intensity of CRs of the $i$-th species 
(in units of $\units{particles~MeV^{-1}~cm^{-2}~sr^{-1}~s^{-1}}$) as a 
function of kinetic energy $T$, the rate $\Gamma_{i}(T)$ of CRs of the $i$-th species
(in units of $\units{particles~MeV^{-1}~s^{-1}}$) impinging on the 
lunar surface will be given by:

\begin{equation}
\Gamma_{i}(T) = 4 \pi R_{\leftmoon}^{2} I_{i}(T) \int \cos \theta_{M} d\Omega_{M} =
4 \pi^{2} R_{\leftmoon}^{2} I_{i}(T)
\label{eq:primrate}
\end{equation}
where $R_{\leftmoon}=1737.1\units{km}$ is the radius of the Moon.
In the previous equation we set $d\Omega_{M} = d \cos \theta_{M} d\phi_{M}$,
where $(\theta_{M}, \varphi_{M})$ are the zenith and azimuth 
angles of CR particles with respect to the lunar surface 
($0 < \cos \theta_{M} < 1$ and $0 < \phi_{M} < 2 \pi$). 

The differential gamma-ray luminosity of the Moon $L_{\gamma}(E_{\gamma})$  
(in units of $\units{photons~MeV^{-1}~s^{-1}}$) is given by:

\begin{eqnarray}
L_{\gamma}(E_{\gamma}) = \sum_{i} \int Y_{i}(E_{\gamma} | T) \Gamma_{i}(T) dT \nonumber \\
= 4 \pi^{2} R_{\leftmoon}^{2} \sum_{i} \int Y_{i}(E_{\gamma} | T) I_{i}(T)~dT
\label{eq:luminosity}
\end{eqnarray}
where $Y_{i}(E_{\gamma}|T)$ is the differential gamma-ray yield 
(in units of $\units{photons~particle^{-1}~MeV^{-1}}$), i.e. the
number of photons per unit energy produced by a primary particle of the $i$-th species.
The yields $Y_{i}(E_{\gamma}|T)$ depend on the mechanisms of interactions 
of primary CRs with the lunar surface (regolith) and on its composition.

The differential intensity of gamma rays (in units of 
$\units{photons~MeV^{-1}~cm^{-2}~sr^{-1}~s^{-1}}$) emitted 
from the Moon can be evaluated starting from the differential luminosity and
is given by:
 
\begin{equation}
I_{\gamma} (E_{\gamma}) = \cfrac{L_{\gamma}(E_{\gamma})}{4\pi^{2} R_{\leftmoon}^{2}}
= \sum_{i} \int Y_{i}(E_{\gamma} | T) I_{i}(T)~dT  
\label{eq:gammaintensity}
\end{equation}

The gamma-ray flux observed by a detector at Earth 
(in units of $\units{photons~MeV^{-1}~cm^{-2}~s^{-1}}$)	
can also be evaluated from the differential luminosity and is given by:

\begin{eqnarray}
\label{eq:moonflux}
\phi_{\gamma}(E_{\gamma}) = \cfrac{L_{\gamma}(E_{\gamma})}{4\pi d^{2}} 
= \cfrac{\pi R_{\leftmoon}^{2}}{d^{2}} I_{\gamma}(E_{\gamma}) \nonumber \\
= \cfrac{\pi R_{\leftmoon}^{2}}{d^{2}} \sum_{i} \int Y_{i}(E_{\gamma} | T) I_{i}(T)~dT  
\end{eqnarray}
where $d$ is the distance between the center of the Moon and the detector.
In the case of the Fermi LAT, due to the orbital motions of 
the Moon and of the Fermi satellite around the Earth,
$d$ ranges from about $3.4\times10^{5}\units{km}$ to
$4.1\times10^{5}\units{km}$ (i.e. from about $54 R_{\oplus}$ 
to $64 R_{\oplus}$, where $R_{\oplus}=6378\units{km}$ is the
mean equatorial Earth radius). 

In particular, Eq.~\ref{eq:moonflux} shows that a $10\%$ change
of the distance $d$ corresponds to a $20\%$ change of the flux.
This effect cannot be eliminated from our data analysis because,
due to the limited photon statistics, in order to properly reconstruct 
the fluxes, we need to analyze data samples collected in periods 
of at least a few months, which are longer than the time scales
corresponding to the orbital periods of the Moon 
($\sim 28 \units{days}$) and of the LAT ($\sim 1.5\units{hours}$).

\section{Data selection}
\label{sec:instrument}

The LAT is a pair conversion gamma-ray telescope,  
sensitive in the energy range from $20 \units{MeV}$ to more than
$300 \units{GeV}$. Here a brief description of the instrument is 
given, while full details can be found in ref.~\cite{atwood2009}.

The instrument is a $4 \times 4$ array of $16$ identical towers,
designed to convert incident gamma rays into $e^{+}e^{-}$ pairs, and
to detemine their arrival directions and energies. Each tower is 
composed of a tracker module and a calorimeter module. 
The tracker consists of $18$ $x-y$ planes of silicon strip detectors 
interleaved with tungsten converter foils, for a total on-axis 
thickness of $1.5$ radiation lengths. 
The calorimeter consists of $96$ CsI (Tl) crystals, hodoscopically arranged 
in $8$ layers. 
The towers are surrounded by a segmented anticoincidence detector
consisting of plastic scintillators, which is used for rejecting
the charged cosmic-ray background.

The analysis presented in this paper has been performed using
the newest {\tt Pass 8} data~\cite{atwood2012}, specifically
P8\_SOURCE photon events starting from a minimum energy of 
$30 \units{MeV}$.

A crucial point in the Moon gamma-ray data analysis is the 
treatment of the background, which originates variously from the diffuse gamma-ray
emission, from the gamma-ray sources that the Moon drifts past along its path
in the sky, and from the tiny residual fraction of charged CRs that are 
misclassified as photons. As the Moon is a moving source, the 
use of a background template might lead to inaccurate results.
Hence, for our analysis we chose to evaluate the background 
directly from the data, by using properly selected signal and background regions. 

The signal region is defined as a cone centered on the Moon 
position, with an energy dependent angular radius given by:

\begin{equation}
\label{eq:angrad}
\theta = \sqrt{
\left[ \theta_{0} ( E/E_{0} )^{-\delta} \right]^{2} 
+ \theta_{min}^{2} } 
\end{equation}
where $E$ is the photon energy, $E_{0}=100\units{MeV}$, 
$\theta_{min}=1\degrees$, $\theta_{0}=5\degrees$
and $\delta=0.8$. The energy dependence of the angular radius follows
the behavior of the $68\%$ containment radius of the 
LAT point-spread function (PSF)~\cite{psf}. This choice maximizes
the signal-to-noise ratio.
The value of $\theta_{min}$ in Eq.~\ref{eq:angrad}
has been chosen to account for the finite dimension
of the Moon, which is seen from the Earth as an extended source
of $0.25\degrees$ angular radius.
The position of the Moon is obtained from its ephemeris using 
software interfaced to the JPL libraries~\cite{ephem} and
correcting for Fermi orbital parallax.

The background region is a cone of the same angular radius 
as the signal region, centered on a time-offset position of the Moon. 
Since the Moon orbits around the Earth with a period of $\sim 28$ days, 
we chose a time offset of $14$ days (i.e. at a given time, 
the center of the background region is in the position that 
the Moon will take $14$ days later). 
We performed our analysis by splitting the data set in smaller subsamples, 
each of one month duration. 
This means that in a month of $30$ days, the center of the 
background region will take $16$ days to reach the position 
occupied by the Moon at the end of that month. When this 
happens, the center of the background region will be brought 
back to the position taken by the Moon at the beginning of the month 
and, starting from this time, it will move along the path 
described by the Moon during the first $14$ days of the month.
In this way the background region will span the same portion of sky
as the signal region and, since the orbital period of the Moon
is close to one month, the angular separation between the centers
of the signal and background regions will always be close 
to $180\degrees$. 

For the analysis of the signal (and background) region we selected 
the time intervals when the LAT was operating in its standard 
science operation configuration and was outside the South Atlantic Anomaly 
(SAA).
To avoid contamination from the bright limb of the Earth we discarded
the data taken during the times when the angular separation between
a cone of angular radius $\theta_{max}=15\degrees$ centered on the 
Moon\footnote{In the analysis of the background region the Moon position
is replaced with the position of the center of the background region.}  
direction and the zenith direction exceeded $100\degrees$.
We also discarded data taken during the times when the 
Moon was observed with off-axis angles $\theta$ larger than $66.4\degrees$
(i.e. $\cos \theta < 0.4$). 
To mitigate the systematic uncertainties due to the bright diffuse gamma-ray
emission from the Galactic plane, in our analysis we selected only the 
periods where the Moon was at a Galactic latitude 
\mbox{$|b_{\leftmoon}|>20\degrees$}. 
We also required a minimum angular distance
of $20 \degrees$ between the Moon and the Sun and between the Moon and any 
bright\footnote{Here we define ``bright'' a source 
whose gamma-ray flux above $100 \units{MeV}$ is larger 
than $2\times 10^{-7} \units{photons~cm^{-2}~s^{-1}}$.}
celestial source in the 2FGL 
Fermi LAT source catalog~\cite{nolan2012}. 
Since the center of the background region spans the same portion of sky 
as the Moon and the good time intervals for the two regions are chosen in the same way, 
the exposures of the signal and of the background regions are nearly identical.

\begin{figure}[!t]
\begin{center}
\includegraphics[width=0.98\columnwidth]{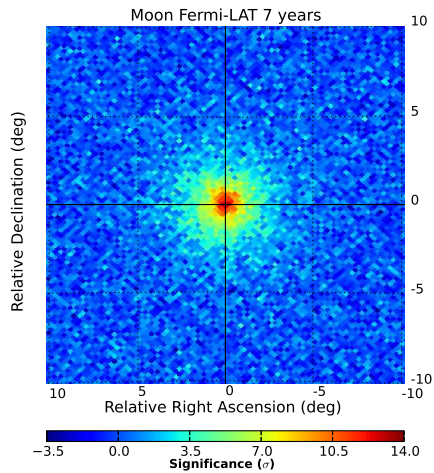}
\end{center}
\caption{Significance map of the Moon as a function of
right ascension and declination relative to the instantaneous Moon
position for
photons in the energy range from $30 \units{MeV}$ to $10 \units{GeV}$. 
The map is built using a HEALPix~\cite{healpix} pixelization of
the sky with $N_{side}=256$ (each pixel corresponds to a solid
angle $\approx 1.6\times 10^{-5} \units{sr}$). The significance
is evaluated following the prescriptions of ref.~\cite{lima}.}
\label{fig:skymap}
\end{figure}

\section{Data analysis and results}
\label{sec:analysis}

Figure~\ref{fig:skymap} shows the significance map of the gamma-ray
signal from the Moon. The map has been built selecting photons
with energies from $30\units{MeV}$ to $10\units{GeV}$.
The significance of each pixel has been
evaluated according to the prescriptions of ref.~\cite{lima},
starting from the counts in the signal and in the background
regions and taking into account the livetime ratio between the
two regions. As expected, the significance map 
exhibits a clear peak in its center, corresponding to
the gamma-ray emission from the Moon. 
The angular size of the peak is broader than that of 
the lunar disk ($0.25\degrees$) due to the finite PSF 
of the LAT and is comparable with the value of the PSF
at $200\units{MeV}$ ($2.9\degrees$), where the peak of the 
signal count spectrum is found. 

\begin{figure}[!t]
\begin{center}
\includegraphics[width=0.98\columnwidth]{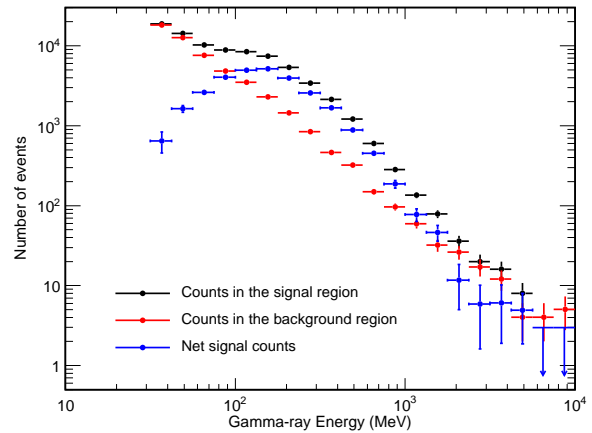}
\end{center}
\caption{Count distributions as a function of 
gamma-ray energy for the signal (black circles) and 
background (red circles) regions. Blue symbols
represent the net signal count spectrum, evaluated by the
method described in ref.~\cite{loparco2011}. 
Circles and associated error bars represent 
the average values and the RMS values of the corresponding PDFs.
Arrows represent upper limits at $95\%$ confidence level.}
\label{fig:counts}
\end{figure}

Figure~\ref{fig:counts} shows the observed photon count spectra 
in the signal and background regions, and the net signal count 
spectrum. The latter was calculated by applying in each energy bin
the Bayesian procedure illustrated in ref.~\cite{loparco2011}, 
taking into account the livetimes of the signal and background 
regions and assuming uniform priors for the net signal counts in 
each energy bin. In particular, for each energy bin we evaluated 
the posterior probability density function (PDF) for the signal 
counts. 
The central values of the net signal count spectrum shown in 
Fig.~\ref{fig:counts} represent the average values of the corresponding 
PDFs, while the error bars represent the corresponding RMSs.
In the energy bins where the significance of the net signal counts 
is smaller than $2 \sigma$ upper limits at $95\%$ confidence level are shown.

To reconstruct the energy spectrum of gamma rays from the Moon 
starting from the observed count spectra and taking energy dispersion
into account, we have implemented an 
analysis method based on the software toolkit BAT~\cite{caldwell2009}. 
The BAT package allows evaluating the full posterior probability 
PDFs for the parameters of a model. 
It is based on Bayes' theorem and is realized with the use of a 
Markov Chain Monte Carlo (MCMC) analysis. In the present work we used BAT to extract, 
starting from the observed count distributions in the signal and background 
regions, the posterior PDFs for both the signal
and background gamma-ray fluxes.

Indicating with $\mu_{s}(E_{i})$ and $\mu_{b}(E_{i})$ the expected
counts in the $i$-th energy bin, respectively in the signal and in 
the background region, it is possible to write the following equations:

\begin{equation}
\mu_{s}(E_{i}) = \sum_{j} P_{s}(E_{i}|E_{j})   
\left[ \phi_{s}(E_{j}) + \phi_{b}(E_{j}) \right] 
A ~ t_{s} ~ \Delta E_{j}
\label{eq:sigflux}
\end{equation}

\begin{equation}
\mu_{b}(E_{i}) = \sum_{j} P_{b}(E_{i}|E_{j}) 
\phi_{b}(E_{j}) ~ A ~ t_{b} ~ \Delta E_{j}.
\label{eq:bkgflux}
\end{equation}
In the previous equations $\phi_{s}(E_{j})$ and $\phi_{b}(E_{j})$ 
are the true signal and background fluxes in the $j$-th energy bin
($\phi_{s}(E)$ corresponds to $\phi_{\gamma}(E)$ in Eq.~\ref{eq:moonflux}),
that are treated as unknown parameters;
$P_{s}(E_{i}|E_{j})$ and $P_{b}(E_{i}|E_{j})$ are the smearing
matrices in the signal and background regions respectively,
i.e. the probabilities that a photon of energy $E_{j}$ is 
observed with energy $E_{i}$, and are evaluated from a full
Monte Carlo simulation of the instrument, taking into account 
the pointing histories of the two regions; $A=6\units{m^{2}}$ is the 
cross sectional area of the spherical surface used for the generation 
of the events in the Monte Carlo simulation; $t_{s}$ and $t_{b}$ are the live times
of the signal and background regions respectively.

If $n_{s}(E_{i})$ and $n_{b}(E_{i})$ are the actual
values of the counts in the $i$-th energy bin of the signal and of
the background regions, it is possible to define the likelihood 
function as a product of Poisson PDFs:

\begin{eqnarray}
\label{eq:batlike}
\mathcal{L}(\vec{\phi}_{s} , \vec{\phi}_{b} ; \vec{n}_{s} , \vec{n}_{b}) = 
\prod_{i} e^{-\mu_{s}(E_{i})} \frac{\mu_{s}(E_{i})^{n_{s}(E_{i})}}{n_{s}(E_{i})!} \times \nonumber \\ 
\prod_{i} e^{-\mu_{b}(E_{i})} \frac{\mu_{b}(E_{i})^{n_{b}(E_{i})}}{n_{b}(E_{i})!}
\end{eqnarray}
where we used the vector notation to denote sets of independent
quantities defined in the various energy bins (i.e. 
$\vec{\phi_{s}}=(\phi_{s}(E_{1}),\phi_{s}(E_{2}),\ldots,)$ etc.). 

\begin{figure}[!t]
\begin{center}
\includegraphics[width=0.98\columnwidth]{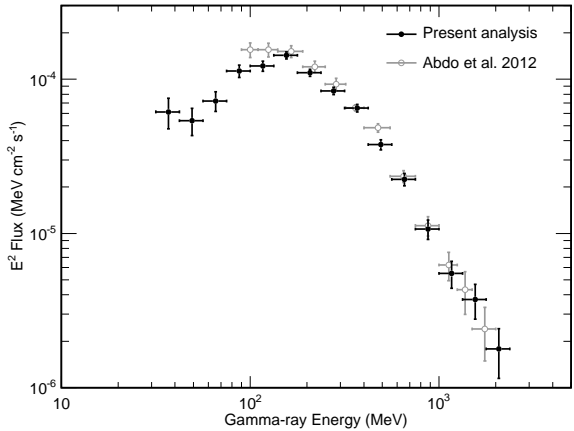}
\end{center}
\caption{Gamma-ray energy spectrum of the Moon.  
The flux values $\phi_{\gamma}(E)$ in each bin are multiplied by $E^{2}=E_{1}E_{2}$, where $E_{1}$
and $E_{2}$ are the lower and upper energy edges of each bin. 
The results from the present analysis (black points) are compared 
with those published in ref.~\cite{abdo2012}. Only statistical error bars are shown.
The central values of each bin represent the mean flux values, while the error bars 
represent the RMSs of the corresponding PDFs.}
\label{fig:fluxes}
\end{figure}

As the starting point for the MCMC we assumed uniform prior PDFs for 
the unknown parameters $\phi_{s}(E_{j})$ and $\phi_{b}(E_{j})$.
The posterior PDFs for $\phi_{s}(E_{j})$ and $\phi_{b}(E_{j})$ are
evaluated by BAT using the likelihood function in Eq.~\ref{eq:batlike}.

Figure~\ref{fig:fluxes} shows the reconstructed gamma-ray spectrum of the Moon.
The present results are compared with those published in 
ref.~\cite{abdo2012}, obtained from the analysis of the first $2$ years 
of data taken by the Fermi LAT. The points shown in the plot correspond 
to the mean values of the PDFs on the signal fluxes in each bin, while the 
error bars indicate the RMS values. The spectral energy distribution $E^{2}\phi_{\gamma}(E)$ is
peaked at about $150\units{MeV}$ and then drops with increasing energy as a power law
with spectral index of about $-2$. 

The present results are consistent with those of ref.~\cite{abdo2012} at
energies above $150\units{MeV}$. The minor discrepancies in the range below
$150\units{MeV}$ can be ascribed to the solar modulation effect on CRs,
which affects the energy spectrum of gamma rays emitted from the Moon
(see also the discussion in sec.~\ref{sec:timeevolution}).
The 2-year interval analyzed in ref.~\cite{abdo2012}
corresponded to the minimum of solar activity at the 
beginning of Solar Cycle 24. On the other hand, 
the dataset used in this analysis spans a period of $7$ years, 
covering more than half of Cycle 24. As a sanity check we applied the analysis
technique illustrated in this paper to the data taken by the LAT 
in the first $2$ years, and the results were consistent with those 
of ref.~\cite{abdo2012} in the whole energy range.

In Fig.~\ref{fig:fluxes} only statistical error bars on the fluxes are shown.
The systematic uncertainties, not shown in Fig.~\ref{fig:fluxes},
are primarily due to the uncertainties on the effective area of the instrument,
which propagate to the gamma-ray fluxes. The uncertainties on the effective area 
were evaluated by the Fermi LAT Collaboration~\cite{systematics}: they drop from
$10\%$ to $3\%$ in the energy range from $30\units{MeV}$ to $100\units{MeV}$ 
and are $\sim 3\%$ at energies above $100 \units{MeV}$. 
Systematic uncertainties are smaller than statistical ones in the whole
energy range: in fact the latter are of $\sim 25\%$ at $30\units{MeV}$,
drop to $\sim 5\%$ at $150\units{MeV}$ and increase again 
to $\sim 25\%$ at $1.5\units{GeV}$.

To search for possible issues in the analysis, in addition to the approach discussed above
and based on BAT, we implemented two more analysis techniques, and we compared the 
results. 

In the first approach, we used the software toolkit MINUIT~\cite{minuit} to evaluate 
the set of parameters $\vec{\phi_{s}}$ and  $\vec{\phi_{b}}$ that maximize the 
likelihood function in Eq.~\ref{eq:batlike}. We find that the results from the MINUIT 
analysis are consistent with those shown in Fig.~\ref{fig:fluxes} within 
the statistical errors in the whole energy range. 

In the second approach, we used an improved version of the bayesian unfolding technique
originally developed by the Fermi LAT Collaboration for the spectral 
analysis of gamma-ray sources~\cite{mazziotta2009,loparco2009,dagostini1995,abdo2010a},
in which we implemented the prescriptions of ref.~\cite{dagostini2010}.  
The starting point for the unfolding procedure is the set of posterior PDFs
for the signal counts in each energy bin, which are used to build a set of 
random realizations of the signal count spectra. These count spectra are 
then unfolded and the corresponding gamma-ray flux spectra are obtained. 
Finally, starting from these spectra, the PDFs on the fluxes in the various 
energy bins are evaluated.
The results from the unfolding analysis are also consistent within the 
statistical errors with those shown in Fig.~\ref{fig:fluxes}.

\begin{figure}[!t]
\begin{center}
\includegraphics[width=0.98\columnwidth]{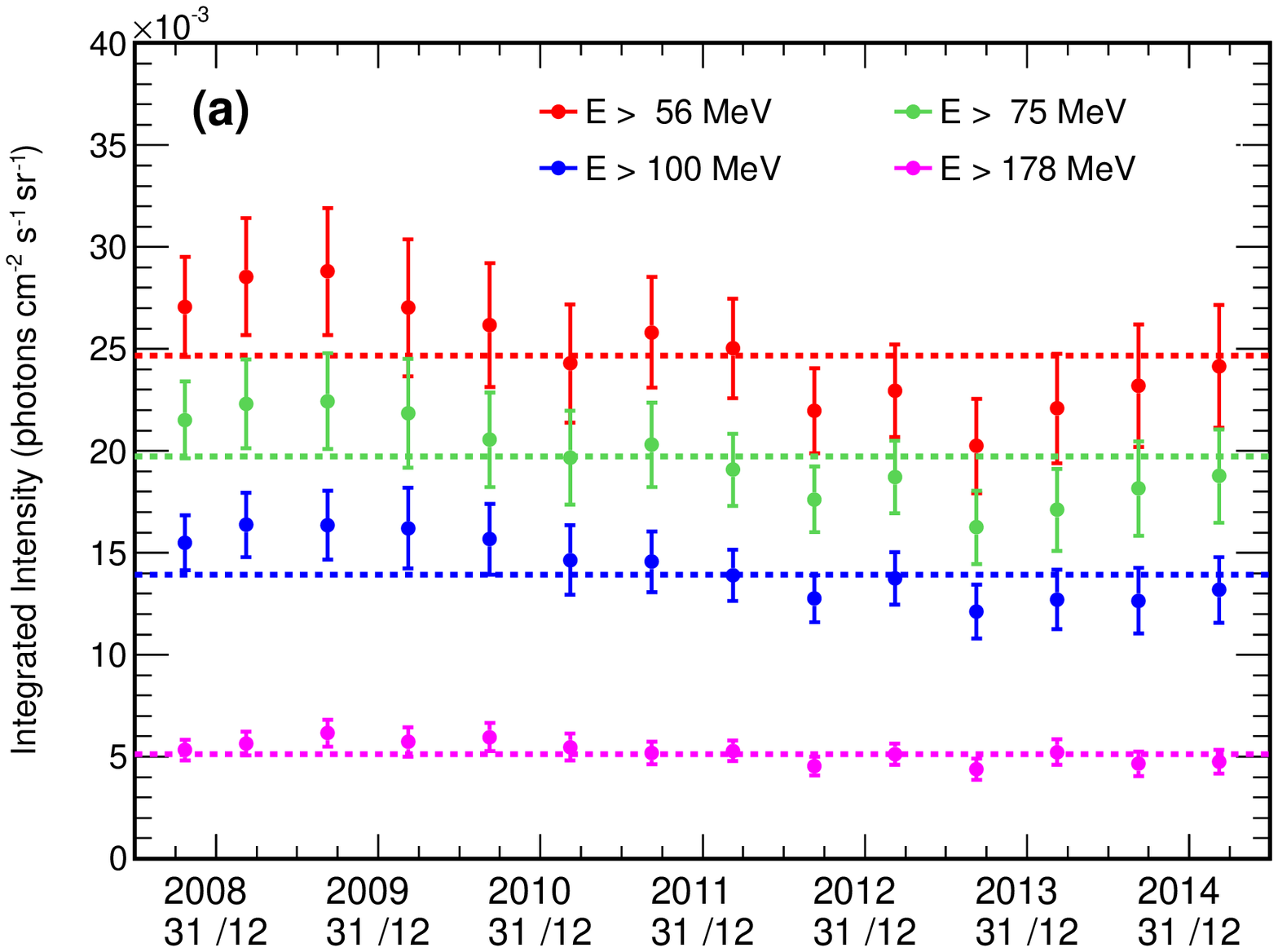}
\includegraphics[width=0.98\columnwidth]{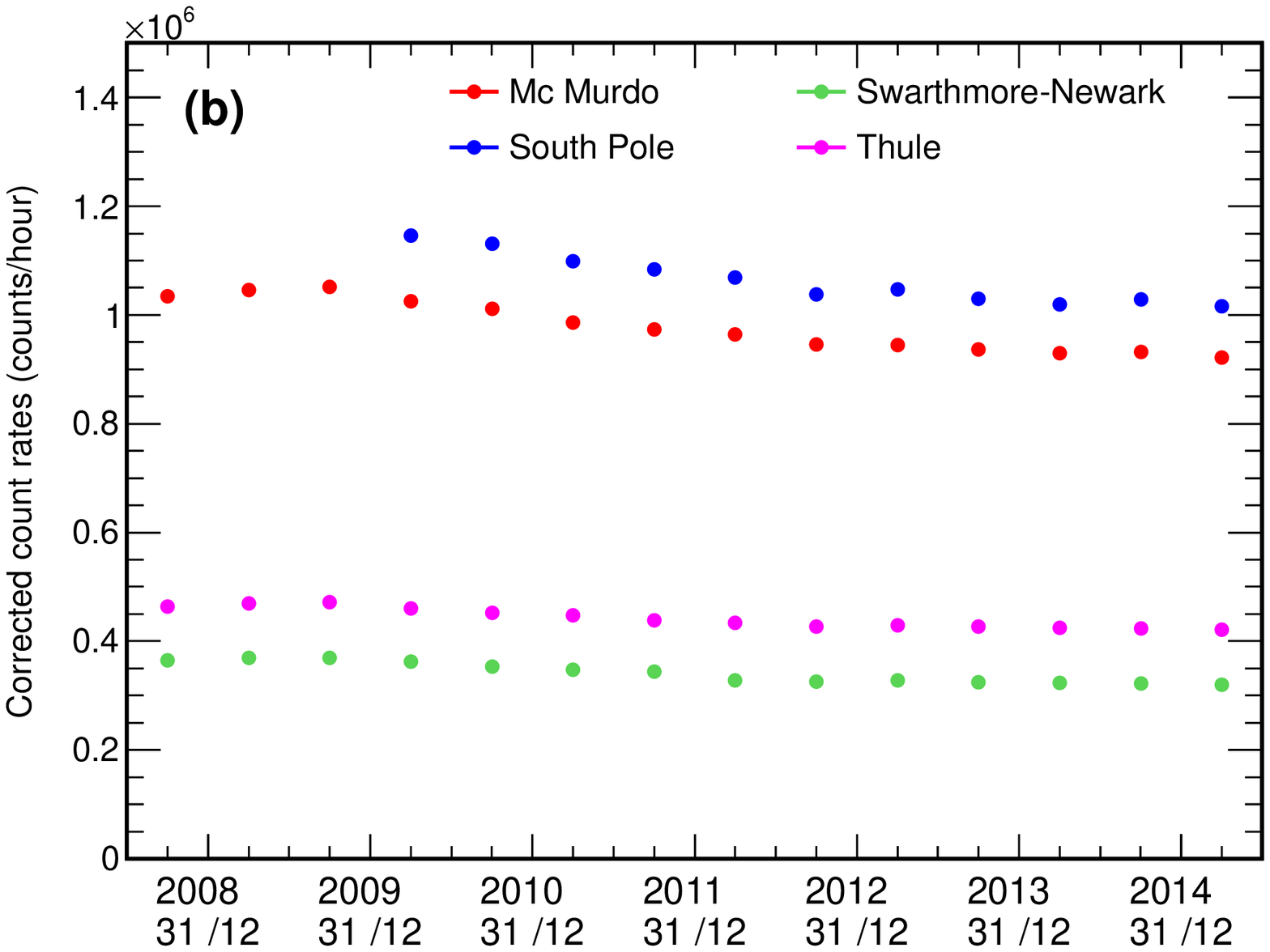}
\end{center}
\caption{(a) Time evolution of the gamma-ray intensity from the Moon.
The red, green, blue and purple symbols represent the intensites above 
$56$, $75$, $100$ and $178\units{MeV}$ respectively. 
The dashed lines indicate the average values
calculated over the whole data-taking period.
(b) Time evolution of the corrected count rates 
registered by the neutron monitors of McMurdo (red), Newark (green), 
South Pole (blue) and Thule (purple). The data of the neutron monitors 
correspond to the good time intervals selected for the Moon data analysis. 
Each point of the plot corresponds to an average value taken over a $6$-month period.}
\label{fig:fluxvstime}
\end{figure}

\section{Time evolution studies}
\label{sec:timeevolution}

To study the time evolution of the gamma-ray emission from the Moon, we
performed the same analysis described in section~\ref{sec:analysis} 
on subsets of data corresponding to $6$-months intervals aligned with 
the beginning of the solar years (i.e. January to June and July to December
except for the first one, starting in August $2008$).

Figure~\ref{fig:fluxvstime}.a shows the time evolution of the 
gamma-ray intensities from the Moon above $56$, $75$, $100$ and $178\units{MeV}$.
The integral intensity is evaluated by integrating the differential
intensity in Eq.~\ref{eq:gammaintensity} over energy. 
The latter is evaluated from the gamma-ray flux using Eq.~\ref{eq:moonflux}.
The error bars shown in the figure have been calculated taking 
into account the statistical uncertainties on the fluxes 
and the variations of the distance between the LAT 
and the Moon during each data-taking period 
(see the discussion in sec.~\ref{sec:formulae}). 
The intensities in the different periods are compared with the
averages, which are calculated considering the whole data-taking period.

Figure~\ref{fig:fluxvstime}.b shows the time evolution of the 
count rates registered by some neutron monitors of the Bartol 
Research Institute~\cite{bartol} installed in various locations in the 
northern (Thule and Newark) and southern (McMurdo and South Pole) 
hemispheres. The count rates are corrected for differences in atmospheric pressure. 
We selected only the neutron monitor data taken during the 
good time intervals selected for the analysis of the Moon 
(see the discussion in sec.~\ref{sec:instrument}). 
The data from the South Pole neutron monitor do not cover 
the whole LAT data-taking period because it was closed
from November 2005 until February 2010.

\begin{figure}[!t]
\begin{center}
\includegraphics[width=0.98\columnwidth]{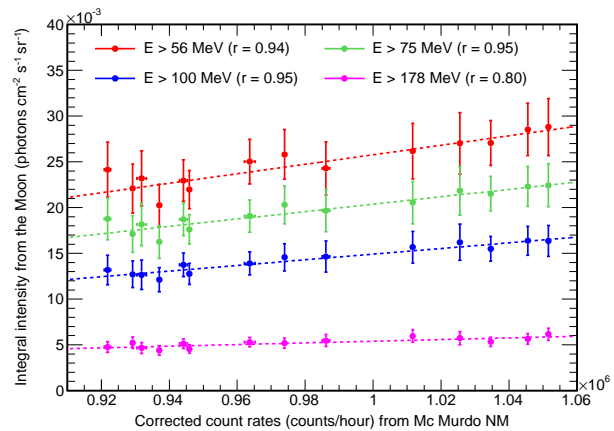}
\end{center}
\caption{Comparison between the gamma-ray integral intensities from the Moon 
above $56$ (red), $75$ (green), $100$ (blue) and $178 \units{MeV}$ (purple) and
the count rate registered by the McMurdo neutron monitor. 
The dashed lines represent the linear regression curves of each series.
The values reported in brackets are the correlation coefficients.}
\label{fig:nmfluxcorr}
\end{figure}

A comparison of the time evolution plots in Fig.~\ref{fig:fluxvstime} suggests that
the gamma-ray emission of the Moon is correlated to the counts 
of the various neutron monitors. 
In Fig.~\ref{fig:nmfluxcorr} we plot the gamma-ray intensities from the Moon
above $56$, $75$, $100$ and $178\units{MeV}$ against the count rates 
registered by the McMurdo neutron monitor. The data indicate that the
lunar gamma-ray emission is indeed correlated with the neutron monitor
count rate. In particular, the correlation is stronger when
the gamma-ray energy threshold is lower and becomes weaker 
as the threshold increases. Similar results are obtained when
comparing the lunar gamma-ray fluxes with the count rates 
registered by other neutron monitors. 
This result is expected, since gamma rays are produced in the 
interactions of primary CRs with the surface of the Moon, and therefore 
their flux must be affected by solar modulation. The correlation
is more evident at low energies, because the solar modulation affects mainly 
the fluxes of low-energy CRs. In particular, in the case of CR protons,
the effect is relevant at kinetic energies $T \apprle 1-10 \units{GeV}$. Since 
the typical energies of gamma rays produced in CR proton interactions are roughly
one order of magnitude less than those of primary protons, the solar modulation
effect is relevant for photons with energies $E_{\gamma} \apprle 0.1-1 \units{GeV}$.

\section{Monte Carlo simulation of CR interactions with the Moon}
\label{sec:fluka}

We have implemented a full Monte Carlo simulation of 
the interactions of CRs with the surface of
the Moon based on the {\tt FLUKA}~\cite{battistoni2007,ferrari2005,flukaweb} 
simulation code. This simulation has been used to evaluate the yields of
gamma rays produced in these interactions. 

{\tt FLUKA} is a general-purpose Monte Carlo code for the simulation of hadronic 
and electromagnetic interactions. It is used in many applications, and is continuously checked 
using the available data from low-energy nuclear physics, high-energy
accelerator experiments and measurements of particle fluxes in the atmosphere.
Hadronic interactions are treated in {\tt FLUKA} following a theory-driven approach. The general phenomenology 
is obtained from a microscopic description of the interactions between the fundamental constituents 
(quarks and nucleons), appropriate for the different energy ranges. 
Below an energy of a few $\units{GeV}$, the hadron-nucleon interactions model is based on resonance 
production and decay, while for higher energies the Dual Parton Model (DPM) is used. 
The extension from hadron-nucleon to hadron-nucleus interactions is done in the framework of 
the Pre-Equilibrium Approach to Nuclear Thermalization (PEANUT) model~\cite{peanut1,peanut2},
including the Gribov-Glauber multi-collision mechanism 
followed by the pre-equilibrium stage and eventually equilibrium 
processes (evaporation, fission, Fermi break-up and gamma deexcitation). 
In case of nucleus-nucleus interactions (in the present work involving
alpha projectiles) {\tt DPMJET-III}~\cite{dpmjet} and a modified version~\cite{ander2004} 
of {\tt RQMD}~\cite{Sor89,Sor89a,Sor95} are used as external event generators,
depending on the projectile energy. More details 
about the FLUKA package can be found in the manual~\cite{ferrari2005,flukaweb} and 
a description of hadronic interaction models used in FLUKA can be found in ref.~\cite{ferrari1996}.

We have calculated the gamma-ray yields from the Moon assuming two different composition models
for the lunar surface. To test these models we have used the Moon gamma-ray 
data taken in the same period as the AMS-02 proton and helium data~\cite{ams02,amshe}. 
We have folded the CR proton and helium spectra measured by AMS-02 with the gamma-ray yields 
predicted by the simulation, and we have compared the resulting predicted fluxes with the data.
Having found good agreement between the model and the data for one of the surface composition models, 
we have assumed a model for the local interstellar spectra (LIS) 
of CR protons and helium nuclei and, starting from the Moon gamma-ray data, we have evaluated 
the solar modulation potential in the framework of the force field approximation.

\subsection{Evaluation of the gamma-ray yield from the Moon}
\label{sec:yield}

As mentioned in sec.~\ref{sec:intro}, in any calculation of the lunar 
gamma-ray emission a Moon surface model must be assumed, which includes a
description of its geometry and its chemical composition.
Regarding the geometry, in our simulation we made the simplest assumption 
that the Moon is a perfect sphere of radius $R_{\leftmoon}=1737.1 \units{km}$, 
thus neglecting the roughness of the lunar surface (the top of the 
highest mountain and the bottom of the deepest crater are within 
$\pm 10\units{km}$ from the surface) and its eccentricity (the
difference between the equatorial radius and the polar radius is 
$< 3 \units{km}$). 

About the chemical composition, we note that the available data are  
from actual samples of lunar rock taken by the 
astronauts in the different landing sites of the Apollo missions and from 
the low-energy gamma-ray, alpha and neutron spectroscopy data~\cite{apollo}. Over 
the years, many models of the lunar surface have been proposed. In particular, 
for the present work, we adopted the lunar surface models proposed by 
Moskalenko and Porter in 2007~\cite{moskalenko2007} (which was also used
in ref.~\cite{abdo2012}) and by Turkevich in 1973~\cite{turk}
(hereafter these models will be indicated in the text as ``MP'' and ``TUR'').
The features of the MP and TUR models are summarized in tab.~\ref{tab:models}. 
The main differences between the two 
models can be found in the weight fractions of the different oxides and 
in the density of the lunar surface. The differences 
result in a lighter composition (lower average atomic and mass numbers) 
of the TUR model with respect to the MP model.

\begin{table}[t]
  \begin{tabular}{L{2.0cm}  C{2.5cm}  C{2.5cm}}
  \hline 
  Model & Moskalenko \& Porter, 2007 & Turkevich, 1973 \\
  \hline
  \hline
  SiO$_{2}$     & $45.0\%$ & $45.0\%$  \\
  FeO           & $22.0\%$ &  $7.6\%$  \\
  CaO           & $11.0\%$ & $15.5\%$  \\
  Al$_{2}$O$_{3}$ & $10.0\%$ & $22.2\%$ \\
  MgO           &  $9.0\%$ &  $8.0\%$  \\
  TiO$_{2}$     &  $3.0\%$ &  $1.1\%$  \\
  Na$_{2}$O     &      $-$ &  $0.6\%$  \\
\hline
  $\rho (\units{g/cm^{3}})$ & $1.80$ & $3.01$  \\
  $\langle Z \rangle$              	 &  $11.5$ &  $10.8$ \\
  $\langle A \rangle$ 			 &  $23.4$ &  $21.8$ \\
\hline
  $X_{0} (\units{g/cm^{2}})$             &  $22.4$ &  $24.4$ \\
  $\lambda_{el} (\units{g/cm^{2}})$      &  $84.5$ &  $82.1$ \\
  $\lambda_{inel} (\units{g/cm^{2}})$    & $150.4$ & $148.4$ \\
\hline
 \end{tabular}
\caption{Summary of the main features of the lunar surface composition models implemented 
in the simulation. The first panel shows the weight fractions of the different oxides composing the lunar
surface. The second panel shows the value of mass density and the average values of the 
atomic number and of the mass number. The last panel shows the values of the radiation length 
and of the proton elastic and inelastic scattering lengths.}
  \label{tab:models}
\end{table}

For both models we have evaluated the gamma-ray yield from the Moon 
by simulating protons and $^{4}$He nuclei with different kinetic energies impinging 
isotropically on the lunar surface. The kinetic energies are
taken on a grid of $81$ equally spaced values in logarithmic scale from 
$100\units{MeV/n}$ to $10\units{TeV/n}$. 
The gamma-ray yield from the $i$-th species of CR primaries
(here $i$=p,$^{4}$He) $Y_{i}(E_{\gamma} | T)$ is calculated as:

\begin{equation}
Y_{i}(E_{\gamma} | T) = \frac{N_{\gamma,i}(E_{\gamma} | T)}
{N_{i}(T) \Delta E_{\gamma}}
\label{eq:yield}
\end{equation} 
where $N_{i}(T)$ is the number of primaries of the $i$-th species generated
with kinetic energy $T$ and $N_{\gamma,i}(E_{\gamma}|T)$ is the
number of photons with energies between $E_{\gamma}$ and
$E_{\gamma} + \Delta E_{\gamma}$ produced by the primaries 
of the type $i$ with
energy $T$ and escaping from the surface of the Moon.

\begin{figure}[!t]
\begin{center}
\includegraphics[width=0.98\columnwidth]{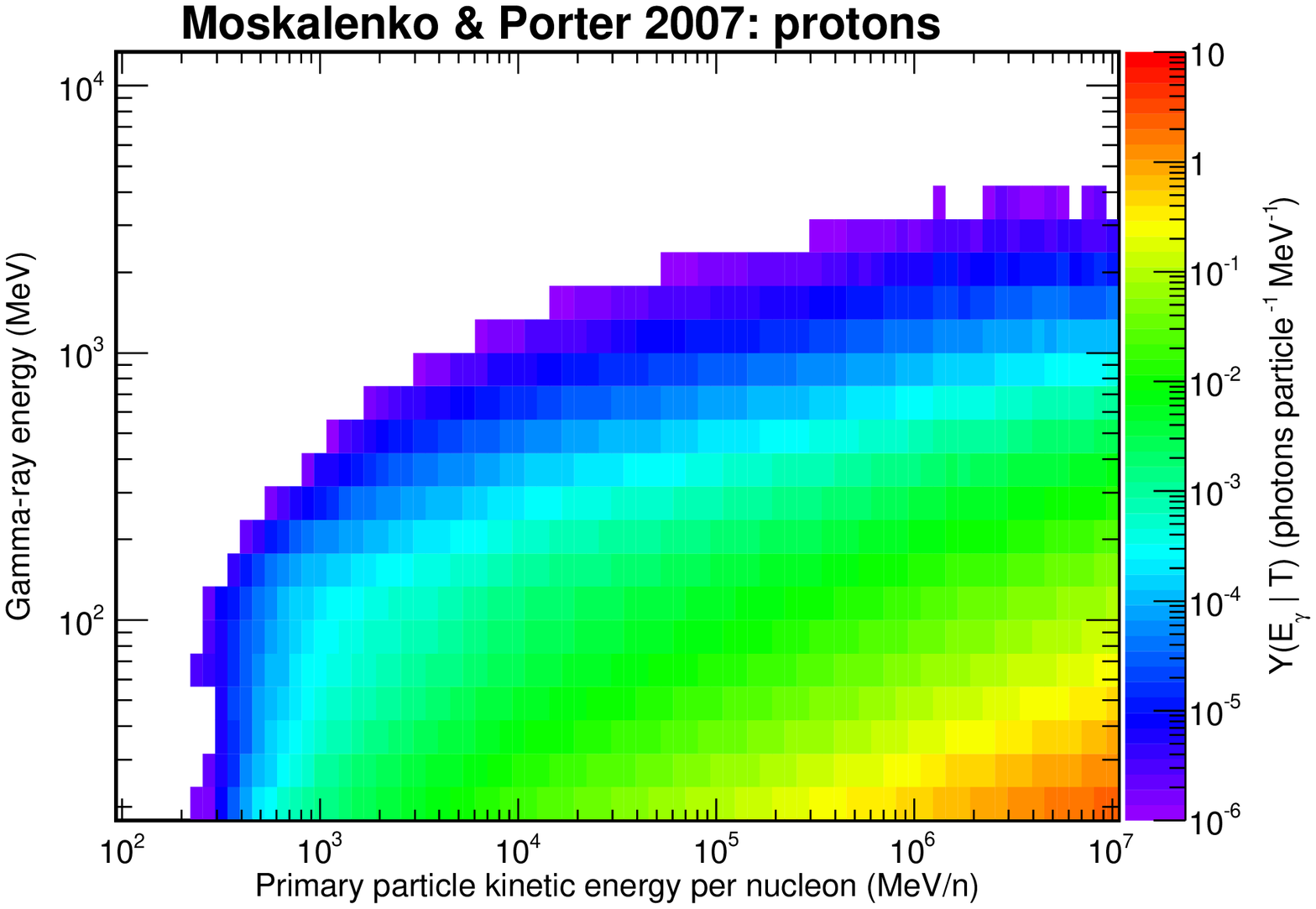}
\includegraphics[width=0.98\columnwidth]{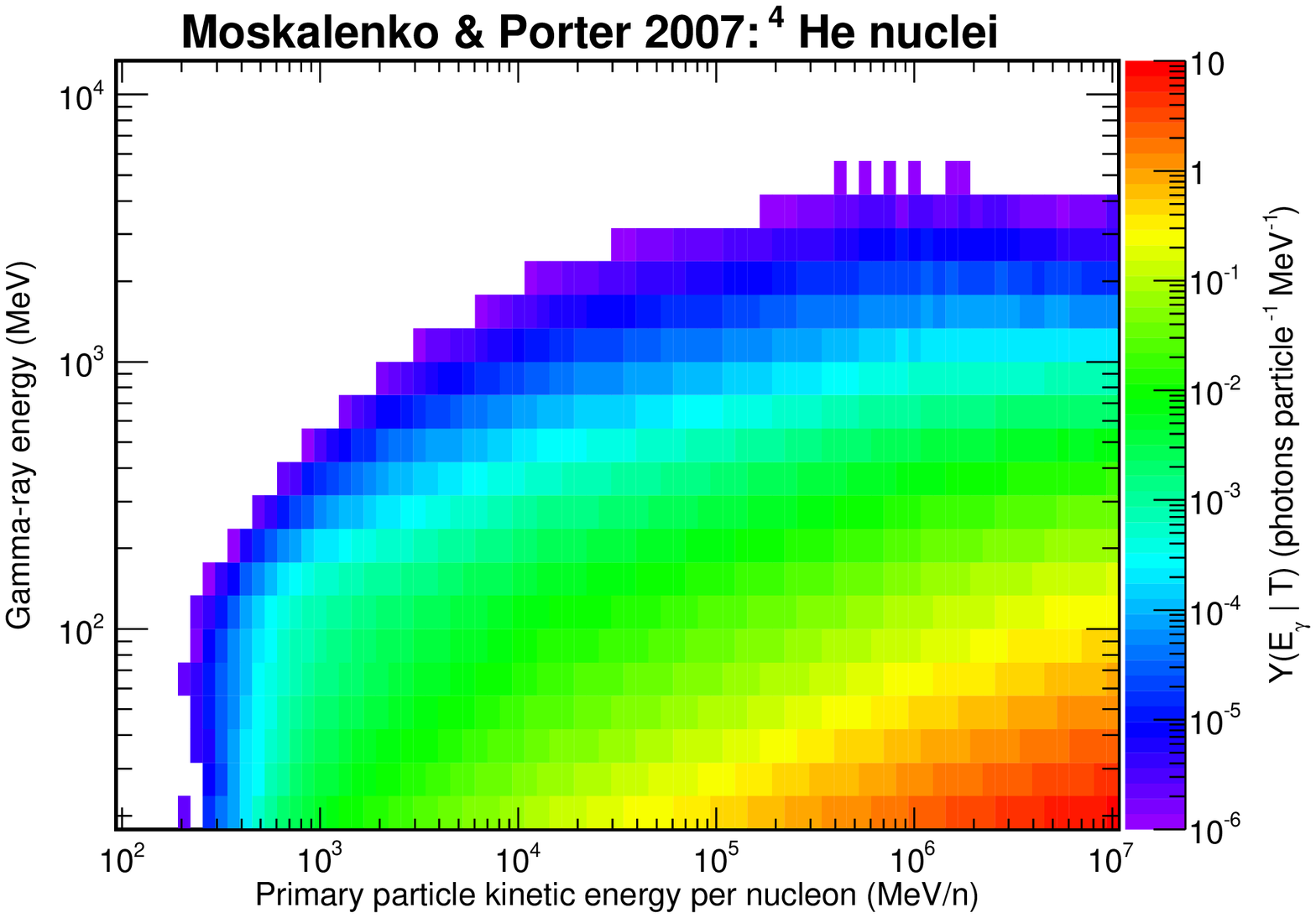}
\end{center}
\caption{Yields of gamma rays produced by the interactions 
of protons (top) and $^{4}$He nuclei (bottom) on the Moon. 
The yields have been evaluated assuming the MP composition model.}
\label{fig:protonyield}
\end{figure}

\begin{figure}[!t]
\begin{center}
\includegraphics[width=0.98\columnwidth]{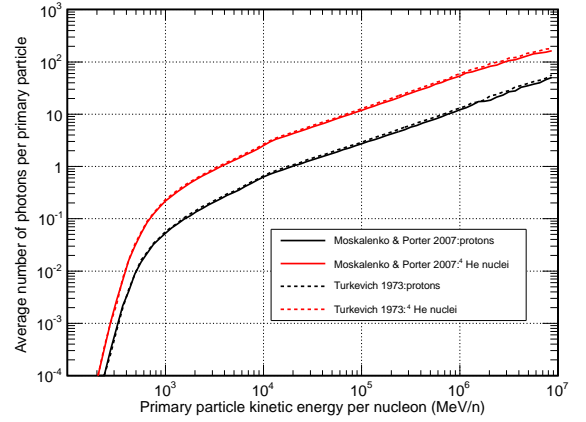}
\end{center}
\caption{Average number of gamma rays per primary particle (in units of $\units{photons/particle}$) 
produced by primary protons (black) and $^{4}$He nuclei (red) as a function of the primary particle
kinetic energy per nucleon. The calculations have been performed for both the MP (continuous lines) 
and the TUR (dashed lines) composition models.}
\label{fig:photperprim}
\end{figure}

Figure~\ref{fig:protonyield} shows the gamma-ray yields 
from the interactions of primary protons and $^{4}$He nuclei
with the Moon calculated with the {\tt FLUKA} simulation as a function of 
the kinetic energy per nucleon of the primary and of the gamma-ray energy 
assuming the MP composition model.
From these plots it is evident that, for both 
proton and $^{4}$He primaries, the gamma-ray yield is negligible
for $T/n \apprle 200~\units{MeV/n}$. This is because 
most gamma rays originate from the decays of neutral pions, and the
process of $\pi^{0}$ production in $p$-nucleus and $^{4}$He-nucleus 
interactions requires a threshold kinetic energy for the incident 
particle. 

\begin{figure}[!t]
\begin{center}
\includegraphics[width=0.98\columnwidth]{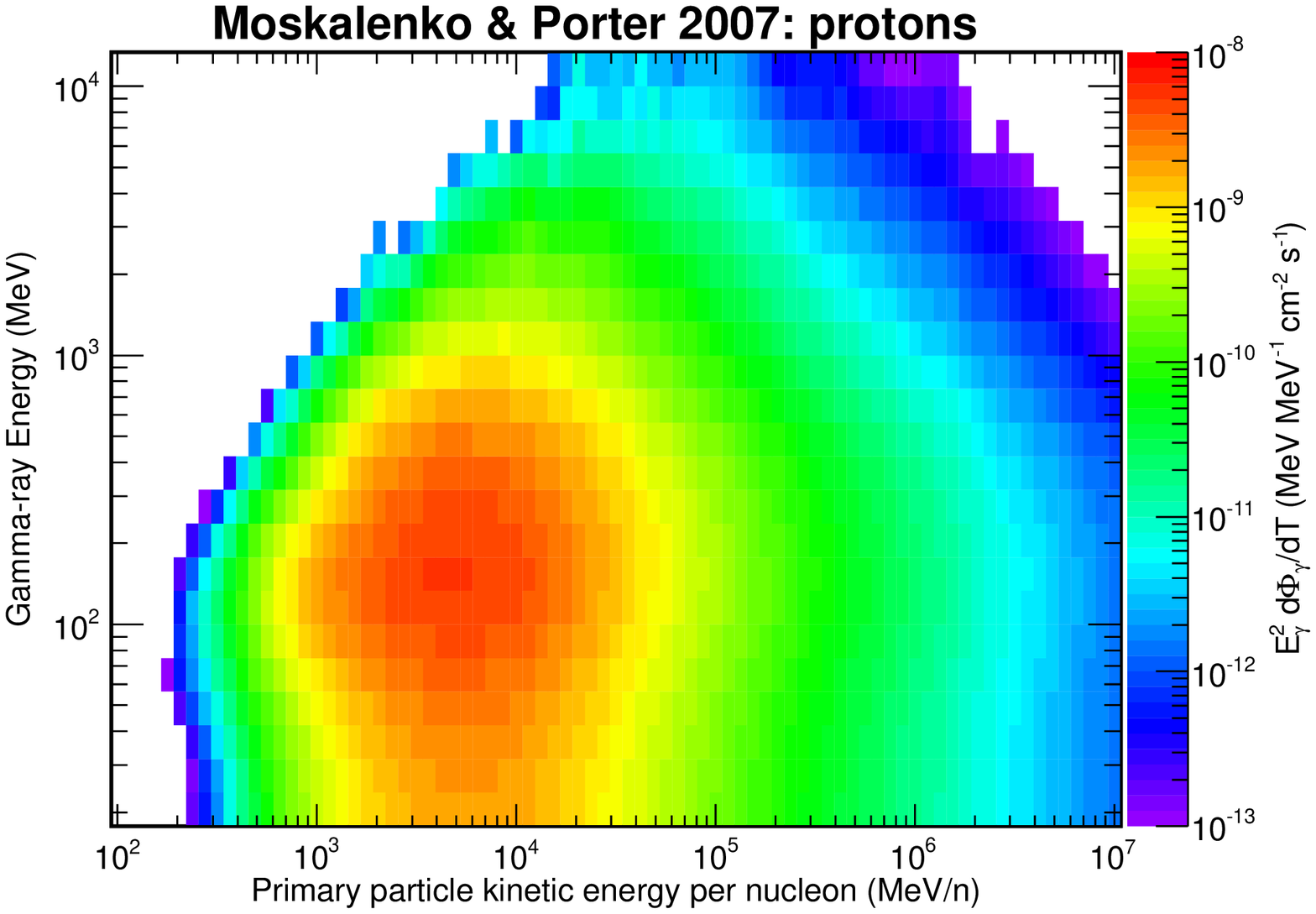}
\includegraphics[width=0.98\columnwidth]{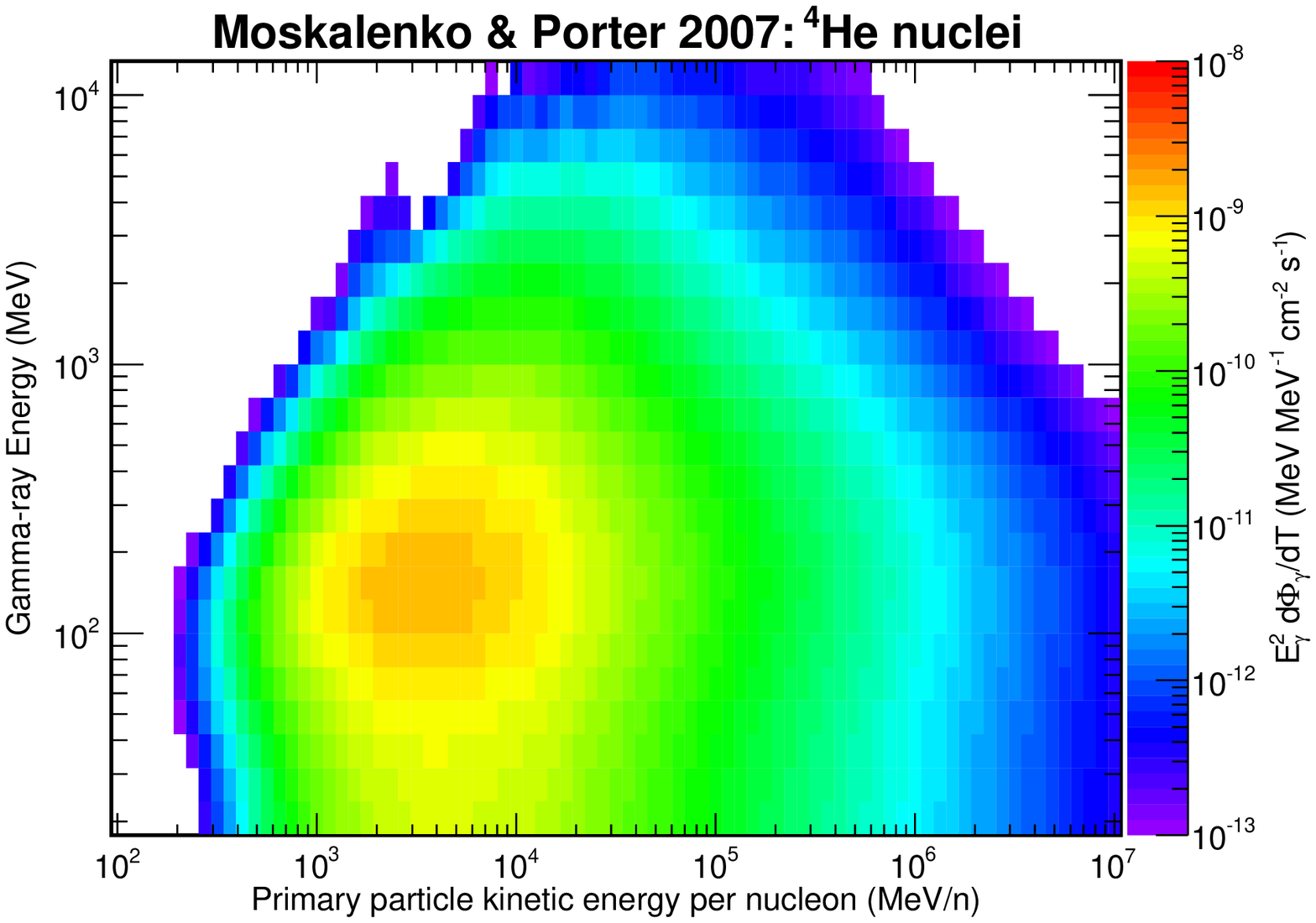}
\end{center}
\caption{Differential photon energy flux from the Moon  
produced by the interactions of protons (top) and $^{4}$He nuclei (bottom) with the Moon surface. 
The photon intensities have been evaluated by folding the gamma-ray yields with the CR proton 
and helium intensity spectra measured by AMS-02~\cite{ams02,amshe}. The calculation has been 
performed with the Moon surface composition model in ref.~\cite{moskalenko2007}.}
\label{fig:foldedyields}
\end{figure}

Figure~\ref{fig:photperprim} shows the average number of photons per primary
particle as a function of the projectile kinetic energy per nucleon produced
by protons and $^{4}$He nuclei, calculated assuming the MP and TUR composition models.
As can be seen in the figure, a $^{4}$He nucleus produces on average 
about four times more gamma rays than a proton with the same kinetic energy per nucleon. 
A simple interpretation of this fact can be given in terms
of the superposition model, according to which a $^{4}$He nucleus is equivalent to
four nucleons.  

Another interesting result is that the gamma-ray yields predicted by the MP and 
TUR models are quite similar. Indeed, a deeper inspection of the results shows 
that the yields calculated with the TUR model are about $20\%$ higher than those 
calculated with the MP model.
The differences could be 
due either to the different compositions or to the different densities. 
To test a possible dependence of the gamma-ray yield on 
the density, we performed some simulations with the TUR and with 
the MP models keeping the composition unchanged and changing 
the density. The results showed that the 
gamma-ray yield is almost independent of the density. We can therefore
conclude that the gamma-ray yield is mainly determined by the chemical
composition of the lunar surface. In particular, the results suggest that
higher values of $\langle Z \rangle$ and $\langle A \rangle$ correspond
to lower gamma-ray yields.

In both these models the lunar surface composition 
is assumed to be independent of depth. Recently, another lunar surface model,
based on the neutron and gamma-ray data from the Lunar Prospector mission,
was proposed by Ota et al.~\cite{ota}, in which the regolith composition and
density are assumed to change with depth. In particular, in the Ota model, 
the lunar surface is described as a stack of four different layers, each
with different thicknesses, compositions and densities (the details of this model 
are given in table 1 of ref.~\cite{ota}).
The gamma-ray yields calculated with the Ota model, not shown in the figure, are 
intermediate between those calculated with the MP and
TUR models. This result was expected, since the values of 
$\langle Z \rangle$ and $\langle A \rangle$ for all the layers composing
the lunar surface are intermediate between those of the MP
and TUR models.

\begin{figure*}[!ht]
\begin{center}
\includegraphics[width=0.98\columnwidth]{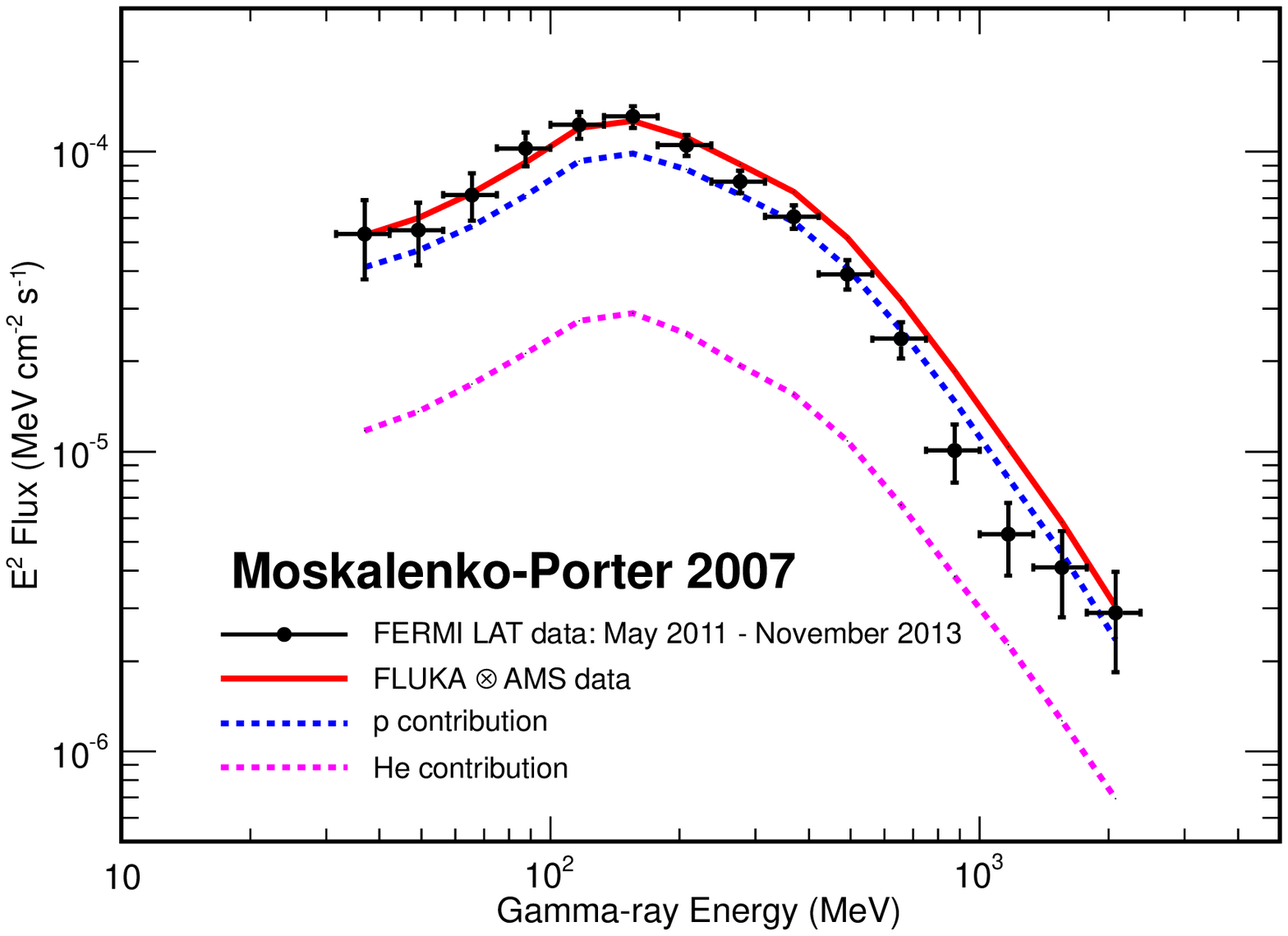}
\includegraphics[width=0.98\columnwidth]{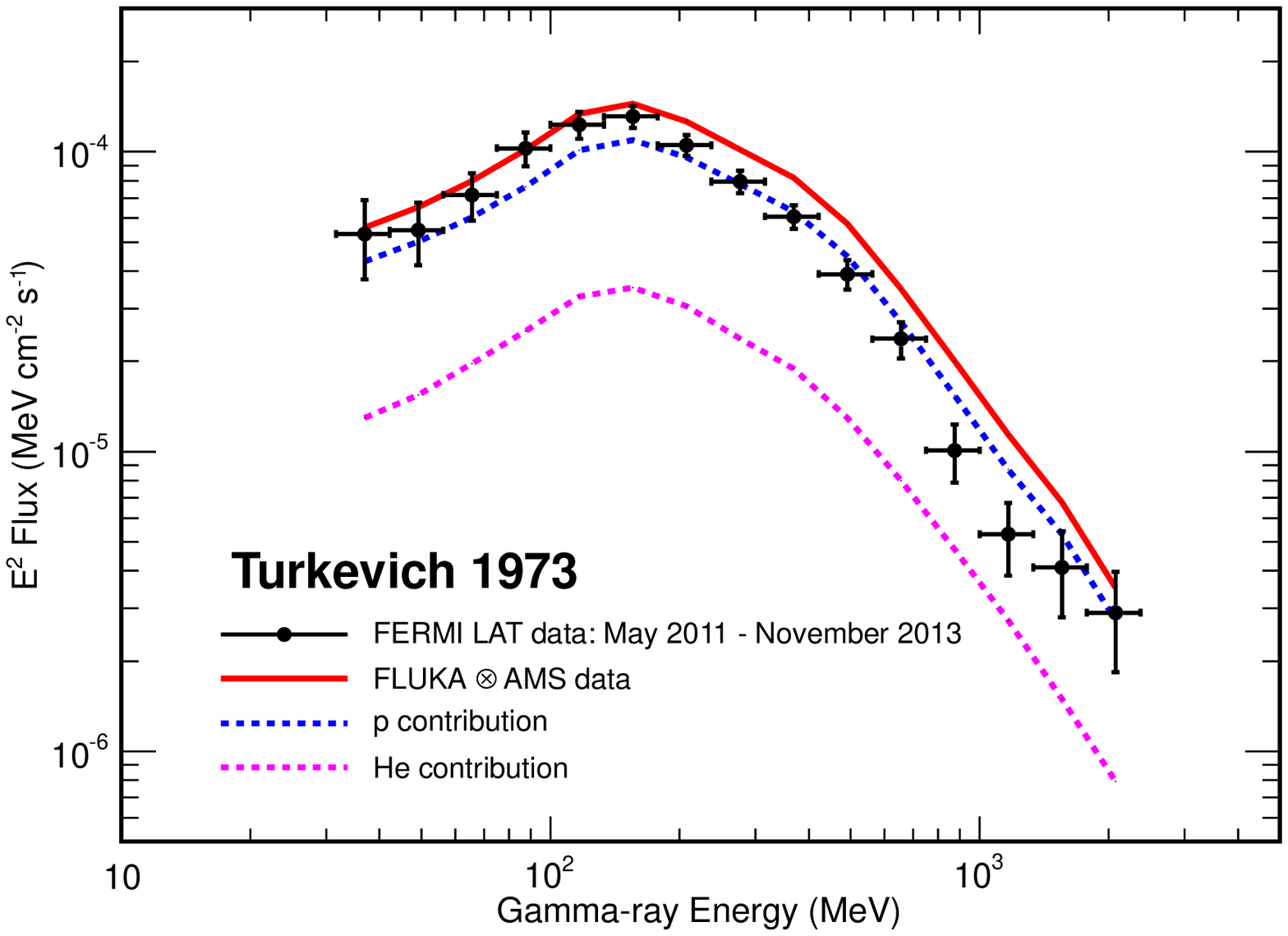}
\end{center}
\caption{Gamma-ray flux from the Moon as a function of energy
in the period May 2011 - November 2013. The results from the LAT data analysis 
(black points) are compared with the expected fluxes obtained 
after folding the CR proton and helium spectra measured 
by AMS-02 in 2011-13 with the gamma-ray yields evaluated in sec.~\ref{sec:yield}
with our simulation. The calculations were perfomed using the lunar surface 
composition models in refs.~\cite{moskalenko2007} (left) 
and~\cite{turk} (right). The continuous red lines indicate the total flux, 
while the dashed blue and purple lines represent the 
contributions to the lunar gamma-ray spectrum from protons and helium nuclei respectively.}
\label{fig:moondirect}
\end{figure*}

\subsection{Evaluation of the lunar gamma-ray spectrum}
\label{sec:spectrum}

To evaluate the lunar gamma-ray intensity spectrum we should fold the spectra of the various
species of CRs impinging on the lunar surface with the gamma-ray yields calculated from the 
Monte Carlo simulation according to Eq.~\ref{eq:gammaintensity}. 
In our calculation we will consider only the contributions from protons and $^{4}$He nuclei,
neglecting those from heavier nuclei. This approximation turns out to be reasonable when 
taking into account the relative abundances of the various CR species. 
Following the considerations in the previous section, we can roughly assume that the  
gamma-ray yields from different nuclei are proportional to the number of their 
constituent nucleons.
Hence, assuming that the relative abundance of CR $^{4}$He nuclei with respect to protons 
is $\sim 10\%$, the contribution of $^{4}$He nuclei to the lunar gamma-ray emission is expected
to be $\sim 40\%$ of the proton contribution, and therefore cannot be neglected. 
On the other hand, if we assume a relative abundance of carbon nuclei with respect to protons 
of $\sim 0.1\%$, we expect their contribution to the lunar gamma-ray emission to be $\sim 1\%$ of 
the proton contribution. Since other CR components are even less abundant than carbon,
we can conclude that the errors from neglecting heavier CR species
in the calculation of the lunar gamma-ray spectrum will be of the order of a few percent.

We also emphasize here that in the calculation of the lunar gamma-ray spectrum 
the isotopic composition of primary CRs should be taken into account. 
However, in the following we will assume that all
CRs with $Z=1$ are protons and all CRs with $Z=2$ are $^{4}$He nuclei. 
Recent measurements~\cite{menn} performed by the
PAMELA experiment show that the $^{2}$H/$^{1}$H ratio decreases from $3.5\%$ to $1.8\%$
in the energy range from $0.1\units{GeV/n}$ up to $1 \units{GeV/n}$, while 
the $^{3}$He/$^{4}$He ratio increases from about $8\%$ up to $18\%$ in the same
energy range. Since deuterons and $^{3}$He are secondaries produced in the 
interactions of primary CRs with the interstellar medium, it is reasonable 
to think that their abundances do not increase significantly at higher energies.
Therefore, assuming these values for the isotopic ratios, we expect that the error on the 
lunar gamma-ray spectrum calculated neglecting the isotopic composition of primary 
CRs will be of percent order.

The contribution to the differential gamma-ray intensity of the Moon 
from the $i$-th species of CR projectiles (protons and $^{4}$He nuclei) 
may be calculated as:

\begin{equation}
\frac{dI_{\gamma,i}(E_{\gamma} | T)}{dT} = Y_{i}(E_{\gamma} | T) I_{i}(T).
\end{equation}
The corresponding photon energy flux can be then evaluated as:

\begin{equation}
E_{\gamma}^{2} \frac{d\Phi_{\gamma,i}(E_{\gamma},T)}{dT} = 
E_{\gamma}^{2} \frac{\pi R_{\leftmoon}}{d^{2}} \frac{dI_{\gamma,i}(E_{\gamma} | T)}{dT} 
\end{equation}
Figure~\ref{fig:foldedyields} shows, for the MP lunar composition model, 
the differential gamma-ray energy fluxes originated by proton and $^{4}$He primaries. 
The calculations have been performed by folding the proton and helium
intensity spectra $I_{p}(T)$ and $I_{He}(T)$ measured by AMS-02~\cite{ams02,amshe} 
with the gamma-ray yields calculated with our simulation\footnote{The AMS-02 
helium spectrum includes both $^{4}$He and $^{3}$He nuclei. Once again it should
be emphasized that we are considering the He primaries as consisting entirely 
of $^{4}$He.}. 
The calculations show that, although the gamma-ray yield increases with 
increasing primary energy, the contribution of high-energy primaries
($T>100\units{GeV}$ in the case of protons) to the lunar gamma-ray emission 
is negligible, due to the shape of the primary intensity spectra
(at high energies $I_{p}(T) \sim T^{-2.7}$ and a similar behavior is
observed for helium primaries). On the other hand, the main contribution 
to the lunar gamma-ray emission comes from primaries with energies 
in the range from about $1\units{GeV/n}$ up to a few tens of $\units{GeV/n}$.

\subsection{Comparison of the Moon gamma-ray data with 
the predictions from direct observations of the CR proton spectrum}
\label{sec:comparison}

As mentioned in sec.~\ref{sec:intro}, the dataset used for this analysis was
taken in a period of time overlapping with the data-taking period
of AMS-02~\cite{ams02,amshe}. This provides, for the first time, the possibility 
to test our Monte Carlo simulation against the direct measurements of 
the CR proton and helium spectra performed by AMS-02.
Our dataset is also partially overlapping with the data-taking period of 
PAMELA. However, at present, a test of the simulation against the PAMELA data is not 
possible. Although the PAMELA Collaboration has measured the
CR proton spectra in two different one-month time intervals 
at the end of 2008 and 2009~\cite{pamela}, they did not provide a
measurement of the helium spectra in the same intervals. 

To test our simulation against the AMS-02 data we selected a data sample
taken in the period from May 2011 to November 2013.
However, it is worthwhile to point out here that the time intervals selected for
our analysis of the gamma-ray emission from the Moon most likely do not match 
those used for the AMS-02 data analysis in ref.~\cite{ams02}. In particular, 
when applying the event selection described in sec.~\ref{sec:instrument}, 
we disregarded those time intervals corresponding to transient events, 
such as solar flares, that might be included in the AMS data analysis.

We then folded the CR proton and helium reference spectra with the 
gamma-ray yields obtained from our simulation with the MP and TUR models.
When evaluating the gamma-ray flux we assumed the LAT-Moon distance
equal to its average value during the data-taking period from 
May 2011 to November 2013. In our calculations we did not take
into account the uncertainties on the proton and helium spectra 
measured by AMS-02, which are of about $2\%$ on average~\cite{ams02,amshe}.
   
Figure~\ref{fig:moondirect} compares the measured gamma-ray fluxes
with the calculations from the Monte Carlo simulation for the two composition models.
As shown in the figure, the gamma-ray spectrum obtained from the 
MP composition model reproduces quite well the data in the whole energy
range, with small discrepancies in the region around $1\units{GeV}$,
where the observed flux is smaller than predicted. 
On the other hand, the spectrum obtained from the 
TUR composition model seems to slightly overestimate the data
in the energy range above $200\units{MeV}$.
According to the discussion in sec.~\ref{sec:yield}, this result can
be attributed to the relatively lighter regolith 
(lower $\langle Z \rangle$ and $\langle A \rangle$)
in the TUR model and the consequently greater gamma-ray yield.

We remark here that, when comparing the data 
with the model predictions, one should also take into account 
all the uncertainties, such as those originating from the fluctuations 
on the LAT-Moon distance (see sec.~\ref{sec:formulae}), those on the 
instrument effective area (see sec.~\ref{sec:analysis}), those on the 
AMS proton and helium spectra (see discussion above) and those on the
hadronic interactions models. All these uncertainties are likely 
of $10\%$ or less.

\begin{figure*}[!t]
\begin{center}
\includegraphics[width=0.98\columnwidth]{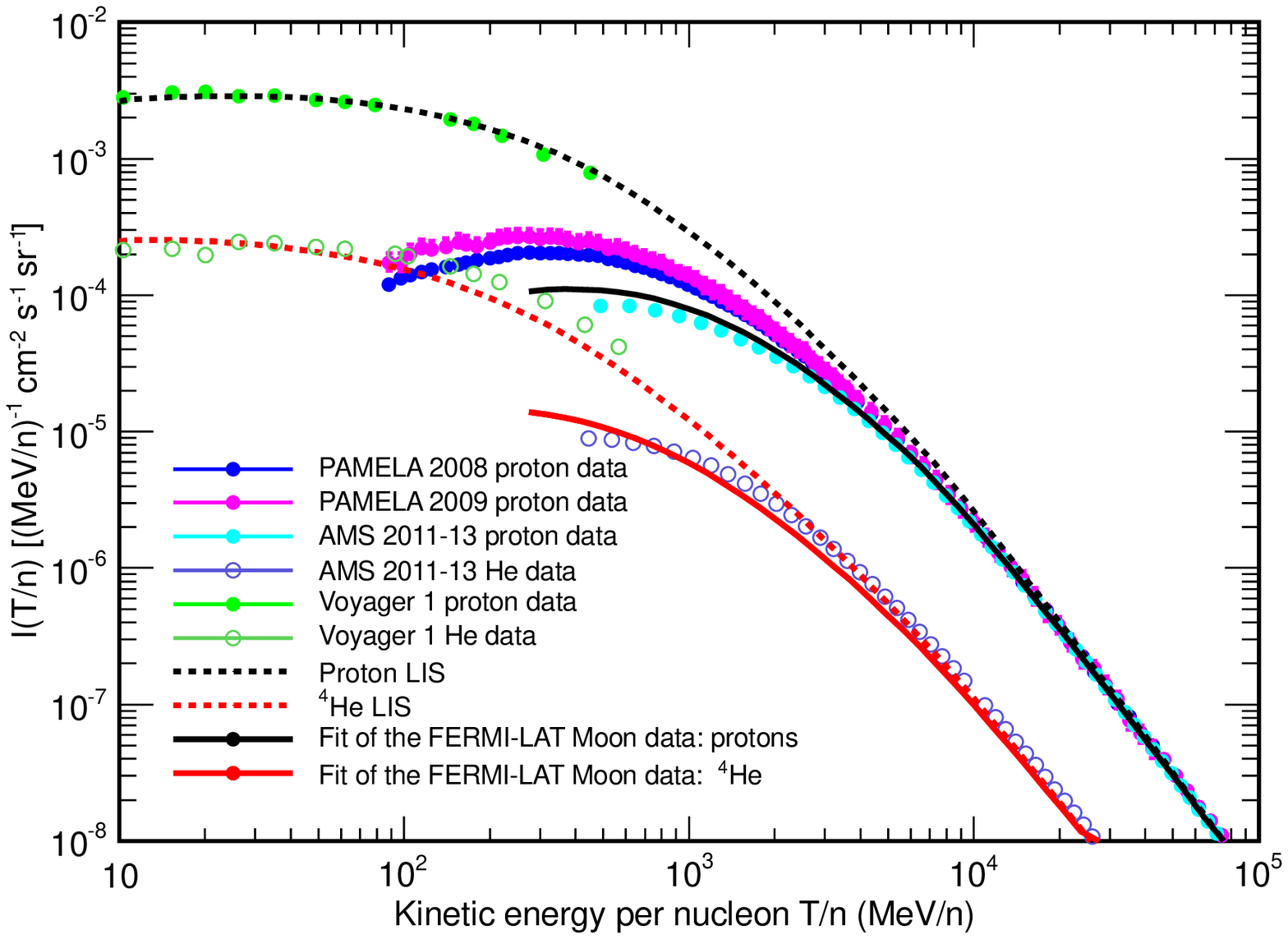}
\includegraphics[width=0.98\columnwidth]{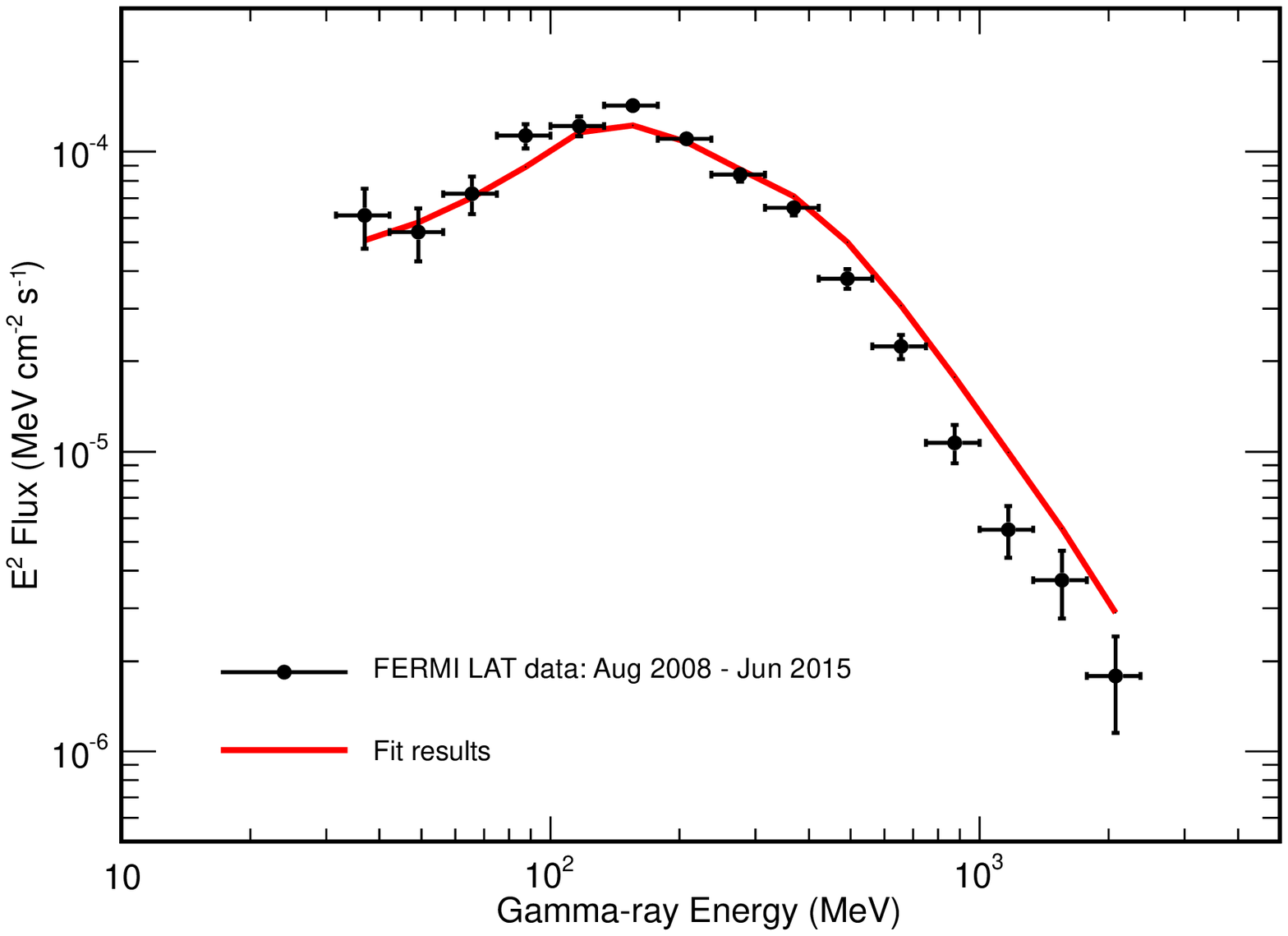}
\end{center}
\caption{Left panel: CR proton and helium spectra obtained from the best fit 
of the Fermi LAT Moon gamma-ray data. The fit was performed using the 
MP lunar surface model. The results of the fit (continuous black and red lines) 
are compared with the proton measurements taken by PAMELA~\cite{pamela} in 
2008 (blue points) and 2009 (purple points) and with the AMS-02~\cite{ams02} 
proton (cyan points) and helium data (violet points). The plot shows also
the proton and helium LIS (dashed black and red lines) 
and the Voyager 1 proton (light green points) and helium (dark green) 
data~\cite{stone2013}. Right panel: Gamma-ray flux from the Moon as a function of energy. 
The results from our analysis are compared with those of the fit. The 
continuous red line represents the average gamma-ray spectrum obtained from the fit, 
assuming that the Moon-LAT distance is equal to its average value during the whole 
data-taking period.}
\label{fig:moonfluka}
\end{figure*}

On the basis of this result, in the following discussion we will adopt 
the MP composition model for the lunar surface.
The small discrepancies between the simulation and the data could be ascribed 
to inaccuracies in our model of CR interactions 
with the Moon. In our model we assume that CR protons of all energies
are impinging isotropically on the whole Moon surface. However, 
low-energy CRs could be affected by the Earth's magnetic field in their
journey to the Moon, in contrast with the hypothesis of an isotropic CR flux.
In addition, in our model we describe the lunar surface as a uniform sphere,
without accounting for the real morphology of the Moon. On the other hand, 
the implementation of a more detailed model would require a huge effort
that is beyond the scope of the present work.

\subsection{Evaluation of the low-energy CR proton and $^{4}$He spectra
and of the solar modulation potential}
\label{sec:flukafit}

The data shown in sec.~\ref{sec:timeevolution} indicate that the lunar 
gamma-ray spectrum is sensitive to the solar modulation effect. 
This is because, as discussed in sec.~\ref{sec:yield}, 
the main contribution to the gamma-ray spectrum of the Moon is that 
of CRs in the energy range up to $\sim 10\units{GeV/n}$. 
In the present section we will illustrate an application of our Monte Carlo 
simulation to the study of the solar modulation potential.

We start from a model for the CR proton and $^{4}$He LIS~\cite{noi,noipaper}, 
evaluated using a customized version of the CR propagation code 
{\tt DRAGON}~\cite{evoli,gaggero}, in which we included a set of 
cross sections for the production of secondary particles in CR 
interactions calculated with {\tt FLUKA}. 
Both the proton and $^{4}$He LIS of ref.~\cite{noi,noipaper} 
were derived in a general framework and, 
together with the LIS of other primary CR components, when propagated to
the solar system, allow to reproduce a wide set of observables.
In particular, these observables include the measurements of CR protons 
performed by PAMELA~\cite{pamela} in 2008 and 2009, the measurements
of CR protons and He nuclei performed by AMS-02~\cite{ams02,amshe} 
from 2011 to 2013, and those performed by Voyager 1~\cite{stone2013} 
during its journey outside the Solar System. 
The proton and $^{4}$He LIS are shown in the left panel of 
Fig.~\ref{fig:moonfluka}, where they are also compared with the data 
from direct measurements. We emphasize here that at high energies 
the $^{4}$He LIS lies below the points measured by AMS-02 because, 
as mentioned in sec.~\ref{sec:spectrum}, the AMS-02 data include 
both the $^{4}$He and $^{3}$He component.

In the following analysis the intensity spectra $I_{i}(T)$ of the various 
CR species (protons and $^{4}$He nuclei) in the Solar System 
are evaluated starting from the LIS intensity spectra $I_{i}^{LIS}(T)$ using the 
force field approximation~\cite{gleeson}:

\begin{eqnarray}
I_{i}(T) & = & I_{i}^{LIS}\left(T + e \Phi Z_{i} /A_{i}  \right) \times \nonumber \\  
& & \cfrac{T \left( T + 2 m_{i}\right)}{\left( T + e \Phi Z_{i} /A_{i} \right) \left( T + e \Phi Z_{i} /A_{i} + 2 m_{i} \right)} 
\end{eqnarray}
where $m_{i}$, $Z_{i}$ and $A_{i}$ are the mass, the charge and the number of nucleons
of the $i$-th primary component, $e$ is the absolute value of the electron charge, and $\Phi$ is 
the solar modulation potential, which in the following discussion will be treated 
as a free parameter.

\begin{figure}[!t]
\begin{center}
\includegraphics[width=0.98\columnwidth]{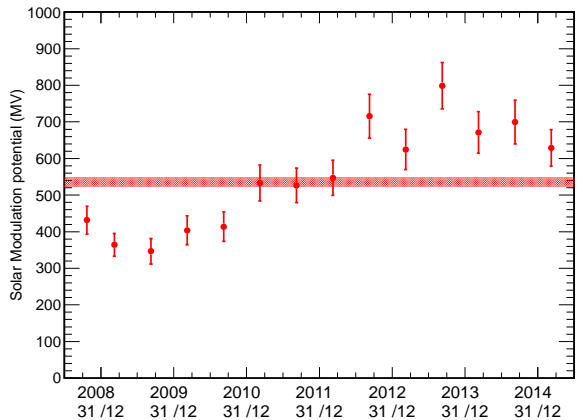}
\end{center}
\caption{Time evolution of the solar modulation potential, evaluated from a fit 
of the lunar gamma-ray emission. The central band corresponds to the average 
value of the solar modulation potential during the whole data-taking period.}
\label{fig:solarmod}
\end{figure}

We used the proton and $^{4}$He LIS and the gamma-ray yields calculated 
with the MP composition model for the lunar surface to perform a fit of the data. 
The fit procedure is based on BAT, and is similar to the one described 
in sec.~\ref{sec:analysis} for the reconstruction of the gamma-ray fluxes
from the Moon. In this case, the gamma-ray signal fluxes in the various 
energy bins are correlated, and are calculated from the cosmic-ray proton 
and helium intensities $I_{p}(T)$ and $I_{He}(T)$ using 
eqs.~\ref{eq:luminosity} and~\ref{eq:moonflux}.
Here the parameters to be fitted are the background 
photon fluxes $\vec{\phi}_{b}$ and the solar modulation potential $\Phi$. 
In our calculations we assumed that the LAT-Moon distance $d$,
that appears in Eq.~\ref{eq:moonflux}, is constant and 
equal to its average value during the whole data-taking period.

The fitting procedure, applied to the whole $7$ year data sample,
yields a solar modulation potential of $537 \pm 12 \units{MV}$. 
The left panel of Fig.~\ref{fig:moonfluka} shows the fitted CR proton 
and helium intensity spectra, compared with the results 
of the direct measurements performed by PAMELA and by AMS-02.
As shown in the figure, the CR proton spectum
inferred from this analysis is consistent with the results from direct 
measurements and lies between the PAMELA and the AMS-02 data.
The helium spectrum lies below the AMS-02 data because, as discussed 
above, it includes only the $^{4}$He component.

The gamma-ray spectrum obtained from the fit is shown in the right panel 
of Fig.~\ref{fig:moonfluka}, where it is compared with the results from 
the data analysis discussed in section~\ref{sec:analysis}. 
The fitted spectrum accurately reproduces 
the data in the energy range up to $400\units{MeV}$, while at higher 
energies it tends to overestimate the measured fluxes.

The fitting procedure discussed here was also applied to the $6$-month data
samples into which the original data set was divided, to study the time evolution
of the solar modulation potential. Fig.~\ref{fig:solarmod} shows the 
time evolution of $\Phi$ obtained from the fit. 
A comparison with the plots in Fig.~\ref{fig:fluxvstime} shows that, as expected,  
the value of the solar modulation potential is 
anticorrelated with the count rates of the various neutron monitors. 
It is also worth noting that, starting in the second half of
2012, the solar modulation potential oscillates about the mean trend 
from interval to interval. This feature
might be due to the major solar flare activity in recent years.

\section{Conclusions}

We measured the fluxes of gamma rays produced by the interactions
of charged CRs impinging on the surface of the Moon using 
data collected by the Fermi LAT from August 2008 to June 2015.
Thanks to the high statistics of the data sample and to the newest 
version of the Fermi LAT event-level analysis and 
instrument response function, we have been able to measure the 
gamma-ray fluxes in an energy range that extends from $30\units{MeV}$ 
up to a few $\units{GeV}$. The time evolution of the flux shows 
that the gamma-ray emissivity of the Moon is correlated 
with the solar activity.

We also developed a full Monte Carlo simulation of the
interactions of CR protons and helium nuclei with the Moon 
using the {\tt FLUKA} simulation code to evaluate the gamma-ray yields.
We implemented two different composition models of the lunar surface 
and we found that the gamma-ray emission from the Moon depends
on the elemental composition of its surface. In particular, we 
observe that the MP composition model provides a good 
agreement between the lunar gamma-ray data and the results of
direct measurements of the CR proton and helium spectra.

Starting from a custom model of the CR proton and helium LIS, we 
then used the simulation to infer the local CR proton intensity
spectrum from the Moon gamma-ray spectrum in the framework of
the force field approximation. The CR spectra obtained 
with this procedure are consistent with the results from direct measurements 
performed by the PAMELA and AMS experiments. We applied this approach to evaluate the time
evolution of the solar modulation potential. The results
show that the potential is anticorrelated with the counts in 
several neutron monitors.

\acknowledgments

The \textit{Fermi} LAT Collaboration acknowledges generous ongoing support
from a number of agencies and institutes that have supported both the
development and the operation of the LAT as well as scientific data analysis.
These include the National Aeronautics and Space Administration and the
Department of Energy in the United States, the Commissariat \`a l'Energie Atomique
and the Centre National de la Recherche Scientifique / Institut National de Physique
Nucl\'eaire et de Physique des Particules in France, the Agenzia Spaziale Italiana
and the Istituto Nazionale di Fisica Nucleare in Italy, the Ministry of Education,
Culture, Sports, Science and Technology (MEXT), High Energy Accelerator Research
Organization (KEK) and Japan Aerospace Exploration Agency (JAXA) in Japan, and
the K.~A.~Wallenberg Foundation, the Swedish Research Council and the
Swedish National Space Board in Sweden.
 
Additional support for science analysis during the operations phase is gratefully 
acknowledged from the Istituto Nazionale di Astrofisica in Italy and the Centre 
National d'\'Etudes Spatiales in France.

The authors acknowledge the use of HEALPix (\url{http://healpix.jpl.nasa.gov/}) 
described in ref.~\cite{healpix}.


\end{document}